\documentclass{IEEEcsmag}

\usepackage[colorlinks,urlcolor=blue,linkcolor=blue,citecolor=blue]{hyperref}
\expandafter\def\expandafter\UrlBreaks\expandafter{\UrlBreaks\do\/\do\*\do\-\do\~\do\'\do\''\do\-}
\usepackage{upmath,color}

\ifCLASSOPTIONcompsoc
  \usepackage[nocompress]{cite}
\else
  \usepackage{cite}
\fi

\usepackage{color, soul}
\usepackage{colortbl}
\usepackage{amsmath}
\usepackage{resizegather}
\usepackage{algorithm}
\usepackage{algorithmic}
\usepackage{tabularx}
\usepackage{graphicx}
\usepackage{array}
\newcolumntype{C}{>{\centering\arraybackslash}p{1cm}}
\usepackage[capitalise]{cleveref}

\newcommand{\nop}[1]{}
\newtheorem{assumption}{Assumption}

\usepackage{booktabs}
\usepackage{tabularx}
\usepackage{booktabs,multirow,array}
\usepackage{subfigure}
\usepackage{url}
\usepackage{float}

\usepackage{footnote}

\usepackage{hyperref}

\begin{document}
%

\sptitle{Theme Article: Data Economy and Data Marketplaces}


\title{Protecting Data Buyer Privacy in Data Markets}

\author{Minxing Zhang}
\affil{Duke University, Durham, NC, 27708, USA, minxing.zhang@duke.edu}

\author{Jian Pei}
\affil{Duke University, Durham, NC, 27708, USA, j.pei@duke.edu}
\markboth{Data Economy and Data Marketplaces}{Data Economy and Data Marketplaces}

\begin{abstract}
Data markets serve as crucial platforms facilitating data discovery, exchange, sharing, and integration among data users and providers. However, the paramount concern of privacy has predominantly centered on protecting privacy of data owners and third parties, neglecting the challenges associated with protecting the privacy of data buyers. In this article, we address this gap by modeling the intricacies of data buyer privacy protection and investigating the delicate balance between privacy and purchase cost. Through comprehensive experimentation, \footnotemark[1] our results yield valuable insights, shedding light on the efficacy and efficiency of our proposed approaches. \footnotemark[2]
\end{abstract}

\maketitle

\footnotetext[1]{Our code is available at \url{https://github.com/minxingzhang0107/Protecting-Privacy-of-Data-Buyers-in-Data-Markets}.}

\footnotetext[2]{The short version of this paper is accepted in IEEE Internet Computing. DOI: \href{https://doi.org/10.1109/MIC.2024.3398626}{10.1109/MIC.2024.3398626}.}
\chapteri{L}arge scale data driven applications rely on more and more data from many different sources.  The power of big data, data science, and AI comes from rich data sources, particularly secondary uses of data. Data markets are online platforms that connect data supplies and demands and enable data discovery, exchange, sharing, and integration~\cite{10.14778/3611540.3611573, 9300226}. Recently, data markets have attracted a lot of interest from both industry and academia.

Privacy holds significant importance in data markets. Broadly defined, it refers to an individual's or a group's capacity to conceal themselves or information about themselves, preventing identification or unwanted approaches by others. In data markets, whether privacy should be treated as goods and how privacy is priced are investigated~\cite{10.1145/2229012.2229054, 10.1145/3219819.3220013}.

Preserving privacy in data markets is of utmost importance. Transactions within data markets have the potential to reveal the privacy of various parties through multiple avenues. Data providers may face privacy risks. 
For instance, hospitals possess valuable medical treatment information that could be sought after by medical equipment companies. If hospitals appropriately collect and anonymize medical treatment data, offering it in data markets while ensuring individual patient identities remain protected, buyers may still glean information, such as the success rates of specific diseases in a hospital, thereby compromising the privacy of the hospital.  Moreover, transactions in data markets may expose the privacy of third parties involved. For instance, an AI technology company providing machine learning model building services to data product buyers could face the risk of model theft~\cite{10.5555/3241094.3241142}, which constitutes a breach of the company's privacy.
To address these privacy concerns in data markets, various approaches are under exploration. 
Strategies include creating decentralized and trustworthy privacy-preserving data markets~\cite{10.14778/3229863.3236266, DBLP:journals/corr/abs-1802-04780}, investigating tradeoffs between payments and accuracy when privacy is a factor~\cite{10.1145/2554797.2554835}, and aggregating non-verifiable information from privacy-sensitive populations~\cite{10.1145/2600057.2602902}.
Numerous studies delve into privacy preservation in data markets, and interested readers are encouraged to explore comprehensive surveys~\cite{10.1145/1749603.1749605, Aggarwal2008, Bertino2008, 10.1007/978-3-540-79228-4_1, 10.1145/1540276.1540279, 7954609, Wu2010} and other sources for further insights. 

While most of the existing studies on privacy preservation in data markets focus on protecting data sellers' and third parties' privacy, data buyers' privacy is often overlooked. Indeed, in data markets, details such as data buyers' identities, purchase locations and times,  products purchased, prices, and quantities can inadvertently expose their privacy. Incidents of e-commerce providers accidentally leaking customer information have been reported again and again, highlighting the urgency of safeguarding buyer privacy.

To the best of our knowledge, there is very limited existing literature that specifically tackles the challenges associated with preserving privacy for buyers in data markets. While Aiello \textit{et al.}~\cite{eurocrypt-2001-2009} tackles the anonymity of buyers in electronic product transactions and introduced an approach aiming at concealing buyer identities during transactions, it is essential to note that, in data markets, the challenges of privacy protection for data buyers remain largely unexplored.

This article addresses the intricate challenge of protecting the privacy of data buyers within a practical and compelling scenario in data markets. In this context, a data buyer publicly declares their purchase intent, specifying the range of the desired data records. This announcement serves to alert potential data owners, encouraging them to provide the relevant data. Simultaneously, the specific details of the purchase intent become the buyer's privacy, requiring protection.

For instance, consider a company planning a financial service targeting customers aged 40-60 with an annual household income of 150-300k, interested in balanced mid-to-long term investment for future retirement. The company wishes to collect financial behavior data from this specific demographic. However, it also aims to keep this strategic focus confidential, preventing competitors from becoming aware of its interest in this customer group. Consequently, the precise range of the data purchase intent must be preserved to ensure the buyer's privacy.

To protect the privacy of data buyers in data markets, we explore a nuanced trade-off between privacy and purchase cost. When a buyer discloses the exact purchase intent, acquiring only the data within that specified intent incurs no additional cost. However, this approach compromises privacy, as an attacker can be certain, with $100\%$ confidence, whether a record is within the buyer's intent or not.

On the other extreme, if a buyer declares the entire data space as the purchase intent, such as customers of any ages and any household income in our previous example, the privacy of the buyer is better preserved. This is because, if the purchase intent represents only a small subset of the entire data space, an attacker has low confidence in determining whether a given data record is part of the buyer's actual intent. However, this heightened privacy comes at a significant cost, as the buyer must purchase all data records in the expansive whole data space.

In a more general approach, buyers can strike a balance between privacy preservation and cost by posting a published intent that is broader than the actual intent, such as customers aged 30-70 and an annual household income 100-500k. 
By doing so, even if an attacker observes the published intent, their confidence in determining whether a data record truly aligns with the exact purchase intent remains below a threshold specified by the buyer. This strategy should minimize the cost associated with purchasing all data records within the published intent. 

In this article, we provide a series of technical contributions. Firstly, we formalize the buyer's privacy preservation problem, considering diverse background knowledge that an attacker may leverage. To the best of our knowledge, we are the first in modeling this specific problem.  Secondly, we introduce heuristic approaches to address the outlined problem. These approaches are designed to enhance the privacy of the buyer while efficiently managing the trade-off with associated costs.  Lastly, we conduct a comprehensive empirical study to evaluate our proposed approach and provide valuable insights into the effectiveness and efficiency of our approach, offering a robust understanding of the practical implications.



\section{Related Work}
As related work, there is a rich body of literature on \emph{privacy-preserving data publishing} (PPDP), which provides methods and tools for publishing useful information while preserving data privacy~\cite{10.1145/1749603.1749605}. 
Most methods for PPDP use some form of data transformation by reducing granularity of representation to protect the privacy~\cite{Aggarwal2008}. The randomization method is a typical PPDP technique by adding noise to the data to mask the attribute values of records~\cite{10.1145/335191.335438}, such as $k$-anonymity~\cite{10.1142/S0218488502001648}, $l$-diversity~\cite{10.1145/1217299.1217302}, and $t$-closeness~\cite{4221659}. Regarding publishing aggregate information about a statistical database, Dwork~\cite{10.1007/11787006_1} proposed the famous $\epsilon$-differential privacy model to ensure that the removal or addition of a single database record does not significantly affect the outcome of any analysis. 
In the distributed privacy preservation setting where an untrusted collector wishes to collect data from a group of respondents, Xue \textit{et al.}~\cite{10.1007/978-3-642-20149-3_9} propose a distributed data collection protocol to allow the data collector to obtain a $k$-anonymized or $l$-diversified version of the aggregated table. 
Recent advancements have introduced data synthesization, which focuses on publishing synthetic datasets that statistically resemble the original data but do not contain any actual user data, thereby mitigating the risk of personal data exposure~\cite{10214943}. To avoid the situation that the output of applications such as query processing, classification, or association rule mining may result in privacy leakage even though the data may not be available, a series of methods were developed to downgrade the effectiveness of applications by either data or application modifications, such as query auditing~\cite{10.1145/76894.76895}, classifier downgrading~\cite{Moskowitz2000ADT}, and association rule hiding~\cite{1269668}.

The predominant focus of the existing studies on PPDP lies in safeguarding the privacy of individual records within a database or across multiple databases. However, the aspect of protecting the privacy of data buyers has not been touched in those studies.

\section{Data Buyer Privacy and Attacker Models}\label{sec:prob}
Consider a data buyer who wants to acquire data in space $\mathcal{D}=D_1 \times \cdots \times D_n$, where $D_1, \ldots, D_n$ are dimensions (also known as attributes).  Denote by $f_\mathcal{D}$ the data distribution in $\mathcal{D}$.  Without loss of generality, we assume that each dimension $D_i$ is finite and nominal.  The target subset of data that the data buyer wants to acquire, called the \textbf{true intent}, can be specified using a conjunctive normal form $V_1 \wedge \cdots \wedge V_n$, where $V_i \subseteq D_i$ $(1 \leq i \leq n)$. When $V_i=D_i$, we also write $V_i$ as  \texttt{ALL}, meaning that the data buyer does not specify any constraint on dimension $D_i$.  Please note that a solution for this basic case may be extended to more general cases, such as an intent like $\vee_{j=1}^m(V_{j,1} \wedge \cdots \wedge V_{j,n})$.  Limited by space, we defer more sophisticated cases to future work.

In an efficient data market, it is assumed that the data buyer specifies the true intent in a truthful manner, that is, all data records falling in the true intent are indeed interesting to the buyer.

For example, consider a dataset representing customer transactions in an e-commerce platform with $3$ dimensions, namely $D_1$, the product category (e.g., electronics, clothing, books), $D_2$, payment method (e.g., credit card, PayPal, cash), and $D_3$, state (e.g., NC, SC, GA).  Now, a data buyer, an online electronics retailer, is interested in acquiring a specific subset of data to analyze customer behavior for targeted marketing. The buyer specifies the true intent truthfully as $(V_1 = \{\text{electronics}\}) \wedge (V_2 = \texttt{ALL}) \wedge (V_3 = \{\text{NC}\})$.

If the specified true intent is released to the public directly, the data buyer's target can be easily obtained by competitors,
revealing a strategic focus on the sale of electronics to customers located in North Carolina. 
This information potentially exposes the buyer to increased competitive pressures, as rivals could leverage this information to refine their marketing strategies, intensify competition in the market, and tailor their approaches to attract customers in the same geographical region.

 A data buyer wants to protect the privacy so that an attacker cannot determine the buyer's true intent with high confidence based on the observable information from the buyer. We will consider different assumptions about the observable information and the corresponding attack models.

To quantify an attacker's intrusion into a buyer's privacy in the true intent, given a record in $\mathcal{D}$, we measure the attacker's confidence about whether the record belongs to the buyer's true intent. The inference about the confidence is part of the attack models to be discussed in the rest of the section.

\subsection{Published Intent-based Attack}
 
To protect the true intent as the buyer's privacy, a data buyer may post a \textbf{published intent} $U_1 \wedge \cdots \wedge U_n$ that is a superset of the true intent, that is, $V_i \subseteq U_i \subseteq D_i$ $(1 \leq i \leq n)$, so that potential data providers know the data buyer's interest.  Our first attack model is based on the published intent (\textbf{PI} for short).

\begin{assumption}[Published intent-based attack]\label{att:pi-uniform}
In a published intent-based attack, an attacker only observes the published intent and does not have any other knowledge about the data space $\mathcal{D}$ and the cost of data collection and processing.  
\end{assumption}

Since the published intent is the only information that an attacker has about the buyer's intent, given a record in $\mathcal{D}$, if the record is outside the published intent, then an attacker knows that the record is not interesting to the buyer. For a record in the published intent, the attacker's confidence on whether the record is in the buyer's true intent can be analyzed as follows. 
 
Since the attacker does not have any background information about the data space, such as the probability density, the best that an attacker can assume is that each possible point in $\mathcal{D}$ has the same probability to occur. That is, the data distribution in $\mathcal{D}$ is uniform.  Therefore, we also call the published intent-based attack the \textbf{PI-uniform attack} in this article.

Since the true intent is a subset of the published intent, and the minimal size of the true intent is $1$, that is, the true intent contains only one specific value on each dimension specified in the published intent.  Thus, the lower bound of the confidence is $\frac{1}{|U_1 \times \cdots \times U_n|}=\prod_{i=1}^n \frac{1}{|U_i|}$, which is known to both the buyer and any attacker. In the same vein, the upper bound of the confidence is $\frac{|V_1 \times \cdots \times V_n|}{|U_1 \times \cdots \times U_n|}=\prod_{i=1}^n \frac{|V_i|}{|U_i|}$, which is only known to the buyer but not to an attacker.  Then, we have
\begin{equation}\label{eq:conf-pi-uniform}
\prod_{i=1}^n \frac{|V_i|}{|U_i|} \geq Conf_{\text{PI}} ^{\text{Uniform}}(x)\geq \prod_{i=1}^n \frac{1}{|U_i|}
\end{equation}
where $Conf_{\text{PI}} ^{\text{Uniform}}(x)$ denotes the attacker's confidence of record $x$ in the data buyer's true intent given $x$ in the published intent via PI-uniform attack.

One may wonder, as $|V_i|$ $(1 \leq i \leq n)$ is unknown to an attacker, how the upper bound $\prod_{i=1}^n \frac{|V_i|}{|U_i|}$ takes effect.  Please note that in the task of buyer privacy protection, the key is to ensure that even an attacker knows the size of the true intent, the chance that a record in the published intent indeed belongs to the true intent is still lower than some buyer specified threshold. Therefore, $|V_i|$ is used in designing the protection mechanisms. 


\subsection{Efficiency Maximization Attack}

In addition to the observable published intent, an attacker may have some background knowledge about the data, such as the data distribution in $\mathcal D$ or the cost of producing or collecting a data record with respect to dimension values.  The attacker may use the background knowledge to enhance the confidence on whether a data record belongs to a data buyer's true intent. The central idea is that, to be economically efficient, a data buyer tends to maximize the efficiency of the published intent.  That is, a large part in the published intent is within the buyer's true intent.  Thus, we call this type of attacks the \textbf{efficiency maximization attacks}.

In our running example, suppose the published intent contains two states, NC and SC.  If an attacker knows that 80\% of the transactions appear in NC, then the attacker may heuristically guess NC is in the true intent.

\begin{assumption}[Efficiency maximization attack]
In an efficiency maximization attack, an attacker observes the published intent and has the background knowledge about the data distribution in the data space $\mathcal D$, i.e. $f_{\mathcal D}$, and/or the cost, i.e. $cost(x)$, of each data record in $x \in \mathcal D$. Here, we assume that $cost(x)$ depends on only the dimension values of $x$.
\end{assumption}

Specifically, if an attacker knows the data distribution in the subspace of the published intent, then, given a record matching the published intent, an upper bound of the attacker's confidence about that the record falls in the buyer's intent is proportional to the probability of the record.  That is,
\begin{equation}\label{eq:conf_upper_bound_only_fd}
Conf_{\text{PI}, f_\mathcal{D}}^{\text{Eff}}(x) \leq \frac{f_\mathcal{D}(x)}{\sum_{t\in U_1 \times \cdots \times U_n}f_\mathcal{D}(t)}
\end{equation}
where $f_\mathcal{D}(x)$ denotes the probability of the record $x$ based on the data distribution in the data space $\mathcal D$.

The above model can be easily extended by incorporating the cost of data records.  By betting that the buyer wants to spend the budget in a cost efficient way, an upper bound of the confidence is 
\begin{equation}\label{eq:conf_upper_bound_both_fd_cost}
Conf_{\text{PI}, f_\mathcal{D}, \text{Cost}}^{\text{Eff}}(x)\leq \frac{f_\mathcal{D}(x)\cdot cost(x)}{\sum_{t\in U_1 \times \cdots \times U_n}f_\mathcal{D}(t)\cdot cost(t)}
\end{equation}

Alternatively, if an attacker knows the cost of data records but does not know the density distribution $f_\mathcal{D}$, still by betting that the buyer wants to spend the budget in a cost efficient way, an upper bound of the confidence is 
\begin{equation}\label{eq:conf_upper_bound_only_cost}
Conf_{\text{PI, Cost}}^{\text{Eff}}(x)\leq \frac{cost(x)}{\sum_{t\in U_1 \times \cdots \times U_n} cost(t)}
\end{equation}


\subsection{Purchased Record Inference Attack}

In some situations, an attacker observes the records purchased by a data buyer, and tries to infer the data buyer's intent. To protect the privacy, a data buyer may purchase more records than the true intent.  Suppose a data buyer purchases a set of records $\mathcal{X} =\{x_1, x_2, \ldots, x_q\}$, containing the records in the true intent and some disguising records that the data buyer uses to protect the privacy. An attacker may try to infer whether the data buyer is interested in a given record. We call this type of attacks the \textbf{purchased record inference attack}.

\begin{assumption}[Purchased record inference attack]
In a purchased record inference attack, an attacker observes the set of data records purchased by the data buyer. Optionally, the attacker may also have the background knowledge about the data distribution in the data space $\mathcal D$, i.e. $f_{\mathcal D}$.
\end{assumption}

Given a data record, an attacker can infer whether the record is interesting to the data buyer in two steps. In the first step, the attacker can infer a pseudo published intent using the observed set of data records purchased by the data buyer. In order words, the attacker can try to find a minimal intent $\hat{U}_1 \wedge \cdots \wedge \hat{U}_n$ such that every purchased data record is in the intent.  That is,
\begin{equation}
  \begin{split}
    \widehat{PI} = & \hat{U}_1 \wedge \cdots \wedge \hat{U}_n \\
     & \min\{\prod_{i=1}^n |\hat{U}_i| \}\\
     s.t.\ & x \subseteq \hat{U}_1 \wedge \cdots \wedge \hat{U}_n \hspace{2mm} \forall x \in \mathcal{X}
  \end{split}\label{pseudo_PI_form}
\end{equation}
Clearly, the solution to Equation~\ref{pseudo_PI_form} can be written as:
\begin{equation}
    \hat{U}_i=\cup_{x \in \mathcal{X}}\{x[i]\}
\end{equation}
where $x[i]$ is the $i$-th dimensional feature of $x$. 

Given a record in $\mathcal{D}$, if the record is outside the pseudo published intent, then an attacker knows that the record is not interesting to the buyer. For a record in the pseudo published intent, the attacker's confidence on whether the record is in the buyer's true intent can be analyzed as follows, using the distribution of purchased records in the pseudo published intent and the attacker's background knowledge.

As in a purchased record inference attack, 
an attacker can infer the observed data distribution $f_{\mathcal P}$ in the purchased set of records, specifically, for a record $x$ in the pseudo published intent, the occurring probability of $x$ in the purchased set can be estimated as
\begin{equation}
    f_{\mathcal{P}}(x) = \frac{h(x)}{|\mathcal{X}|}
\end{equation}
where $h(x)$ denotes the occurring frequency of record $x$ in $\mathcal{X}$, and $|\mathcal{X}|$ denotes the cardinality (total number of records in $\mathcal{X}$).

In the simplest case, the attacker only observes the records purchased by the data buyer and does not have any background knowledge about the data distribution in the data space $\mathcal D$. Then, given a record matching the pseudo published intent, an upper bound of the attacker's confidence about that record falling in the buyer's true intent is proportional to the occurring probability of the record in $\mathcal{X}$. That is, 
\begin{equation}\label{eq: PRI_no_back}
    Conf_{\mathcal{X}}^{\text{Purchased}}(x) \leq f_{\mathcal P}(x)
\end{equation}

If an attacker has the background knowledge about the data distribution in the data space $\mathcal D$, i.e. $f_{\mathcal D}$, by comparing $f_\mathcal{D}$ and the observed distribution $f_\mathcal{P}$, the attacker can infer how likely a purchased record $x$ is in the true intent based on the efficiency maximization assumption.  The more frequent $x$ in the observed distribution $f_\mathcal{P}$ and less likely in $f_\mathcal{D}$, the more likely $x$ belongs to the true intent.

Quantitatively, for a purchased record $x$, we measure how unlikely the frequency of $x$ in the observed distribution $f_\mathcal{P}$ may happen by chance given the distribution $f_\mathcal{D}$, which is indicated by the p-value. We first compute the occurring probability of $x$ in the psedo published intent $\widehat{PI}$ based on $f_\mathcal{D}$ as follows:
\begin{equation}\label{eq:ground_truth_data_distribution_PI}
    f_{\mathcal D, \widehat{PI}}(x) = \frac{f_{\mathcal D}(x)}{\sum_{t\in \hat{U}_1 \times \cdots \times \hat{U}_n}f_\mathcal{D}(t)}
\end{equation}

If $f_\mathcal{P}(x) > f_{\mathcal D, \widehat{PI}}(x)$ and the p-value is small, then likely $x$ belongs to the true intent. To compute its p-value, we conduct a model-less hypothesis testing by leveraging Monte Carlo simulation and permutation testing. More specifically, we sample $\mathcal{L}$ sets of records from the pseudo published intent based on $f_{\mathcal D, \widehat{PI}}$ each with the same size as the buyer's purchased set $\mathcal{X}$ (denoted as $\mathcal{X}_1, ..., \mathcal{X}_\mathcal{L}$). For each record $x$ in $\mathcal{X}$ satisfying $f_\mathcal{P}(x) > f_{\mathcal D, \widehat{PI}}(x)$, we compute the p-value of $f_\mathcal{P}(x)$ (the proportion of sampled sets with record $x$'s occurring probability at least as extreme as $f_\mathcal{P}(x)$). With this, an upper bound of the attacker's confidence about that record falls in the buyer's intent can be written as:
\begin{equation}\label{eq:confidence_upper_bound_PRI_f_d}
    Conf_{\mathcal{X}, f_{\mathcal D}}^{\text{Purchased}}(x) \leq 1 - p\_value(f_\mathcal{P}(x))
\end{equation}

The aforementioned model readily lends itself to extension by integrating the cost of data records, following the same logic as the efficiency maximization attack as indicated in Equation~\ref{eq:conf_upper_bound_both_fd_cost}. For conciseness, we refrain from delving further into the exploration of data record costs in purchased record inference attack.

\subsection{Problem Formulation}

Given a buyer's true intent and a buyer-specified maximal confidence threshold $\lambda>0$, the data buyer wants to either release a published intent in the PI-uniform attack and efficiency maximization attack models or purchase data records in a larger scope in the purchased record inference attack, so that any attackers' confidence is not higher than $\lambda$ on a record in the published intent or purchased by the data buyer belonging to the buyer's true intent. A published intent is called a \textbf{$\lambda$-privacy preserving published intent}, and a set of data records to be purchased is called a \textbf{$\lambda$-privacy preserving set of purchased records} if they satisfying the requirement.

Since a data buyer has to pay for the records in the published intent or the set of data records to be purchased, for the sake of economic consideration, the published intent and the scope of data records to be purchased should be minimized. The \textbf{problem of data buyer privacy protection} is to compute the minimal $\lambda$-privacy preserving published intent or set of purchased records.

\subsubsection{PI-uniform Attack}
In this case, we need to ensure that the confidence upper bound as expressed in Equation~\ref{eq:conf-pi-uniform} is not higher than $\lambda$. Thus, we have
$
Conf_{\text{PI}, \text{Uniform}} \leq \prod_{i=1}^n \frac{|V_i|}{|U_i|} \leq \lambda
$.
That is,
$
\frac{\prod_{i=1}^n |V_i|}{\lambda} \leq \prod_{i=1}^n |U_i|$.
In other words, the defence to a PI-uniform attack can be modeled as the following optimization problem.
\begin{equation}
  \begin{split}
    PI = & U_1 \wedge \cdots \wedge U_n \\
     & \min\{\prod_{i=1}^n |U_i| \}\\
     s.t.\ & \frac{\prod_{i=1}^n |V_i|}{\lambda} \leq \prod_{i=1}^n |U_i|
  \end{split}
\label{PI-uniform_objective}
\end{equation}

\subsubsection{Efficiency Maximization Attack}

In this case, we need to ensure that the confidence upper bound based on an attacker's background knowledge about the data distribution as expressed in Equation~\ref{eq:conf_upper_bound_only_fd} is not higher than $\lambda$ for all records belonging to the true intent (it is also generalizable given an attacker's background knowledge about both the data distribution and the costs, as indicated in Equation~\ref{eq:conf_upper_bound_both_fd_cost}). For any $x \in V_1 \times \cdots \times V_n$,
$
Conf_{\text{PI}, f_\mathcal{D}}^{\text{Eff}}(x) \leq \frac{f_\mathcal{D}(x)}{\sum_{t\in U_1 \times \cdots \times U_n}f_\mathcal{D}(t)} \leq \lambda
$.
We note that,
\begin{equation}
\frac{f_\mathcal{D}(x)}{\sum_{t\in U_1 \times \cdots \times U_n}f_\mathcal{D}(t)} \leq \frac{\max_{r \in V_1 \times \cdots \times V_n}f_{\mathcal{D}}(r)}{\sum_{t\in U_1 \times \cdots \times U_n}f_\mathcal{D}(t)} \nonumber
\end{equation}
That is,
$ \frac{\max_{r \in V_1 \times \cdots \times V_n}f_{\mathcal{D}}(r)}{\lambda} \leq \sum_{t\in U_1 \times \cdots \times U_n}f_\mathcal{D}(t)
$.

In other words, the defence to an efficiency maximization attack can be modeled as the following optimization problem.
\begin{equation}
  \begin{split}
    PI = & U_1 \wedge \cdots \wedge U_n \\
     & \min\{\prod_{i=1}^n |U_i| \}\\
     s.t.\ &  \frac{\max_{r \in V_1 \times \cdots \times V_n}f_{\mathcal{D}}(r)}{\lambda} \leq \sum_{t\in U_1 \times \cdots \times U_n}f_\mathcal{D}(t)
  \end{split} \label{EM_objective}
\end{equation}

\subsubsection{Purchased Record Inference Attack}
In this case, we need to ensure that the confidence upper bound based on an attacker's background knowledge about the data distribution as expressed in Equation~\ref{eq:confidence_upper_bound_PRI_f_d} is not higher than $\lambda$ for all records belonging to the true intent. For any $x \in V_1 \times \cdots \times V_n$,
$$
    Conf_{\mathcal{X}, f_{\mathcal D}}^{\text{Purchased}}(x) \leq 1 - p\_value(f_\mathcal{P}(x)) \leq \lambda
$$
That is,
$$
     1 - \lambda \leq p\_value(f_\mathcal{P}(x))
$$

Thus, we need to make sure that the p-value of $f_\mathcal{P}(x)$ is greater than or equal to $1-\lambda$. Considering utility as a primary consideration, where the buyer seeks to maximize the ratio of the number of records in the true intent against that in the published intent, the defense to a purchased record inference attack can be modeled as the following optimization problem.
\begin{equation}
  \begin{split}
     & \max_{\mathcal{X}\subseteq \mathcal{D}}\Big\{\frac{|\{r\}|}{|\mathcal{X}|}\Big\} \\
     s.t.\ &  1 - \lambda \leq p\_value(f_\mathcal{P}(r)) \quad \forall r \in \mathcal{X}, r \in V_1 \times \cdots \times V_n
  \end{split} \label{PRI_objective}
\end{equation}

\section{Data Buyer Privacy Protection}\label{sec:algo}

In this section, we develop the algorithms tackling the data buyer privacy preservation problem.

\subsection{PI-uniform Attack / Efficiency Maximization Attack}

Given space $\mathcal{D}=D_1 \times \cdots \times D_n$ and the buyer's true intent $V_1 \wedge \cdots \wedge V_n$, we can keep expending the set of feature values $V_i$ in the true intent until obtaining a published intent that can protect the true intent sufficiently, that is, the constraint in Equations~\ref{PI-uniform_objective} (PI-uniform attack) or~\ref{EM_objective} (efficiency maximization attack) is satisfied.

Specifically, we set the initial published intent $PI = U_1 \wedge \cdots \wedge U_n$ to the same as the true intent, that is, $U_i=V_i$ for $1 \leq i \leq n$. In each iteration, we iterate through every dimension in $\mathcal{D}$; for dimension $D_i \in \mathcal{D}$ and the corresponding feature value set $U_i$ in the published intent, we select the nearest neighbor feature value in $D_i \setminus U_i$. After iterating through every dimension, we add the best neighbor feature value to the published intent that minimizes Equations~\ref{PI-uniform_objective} (PI-uniform attack) or~\ref{EM_objective} (efficiency maximization attack). 

\begin{figure}[t]
\includegraphics[width=0.5\textwidth]{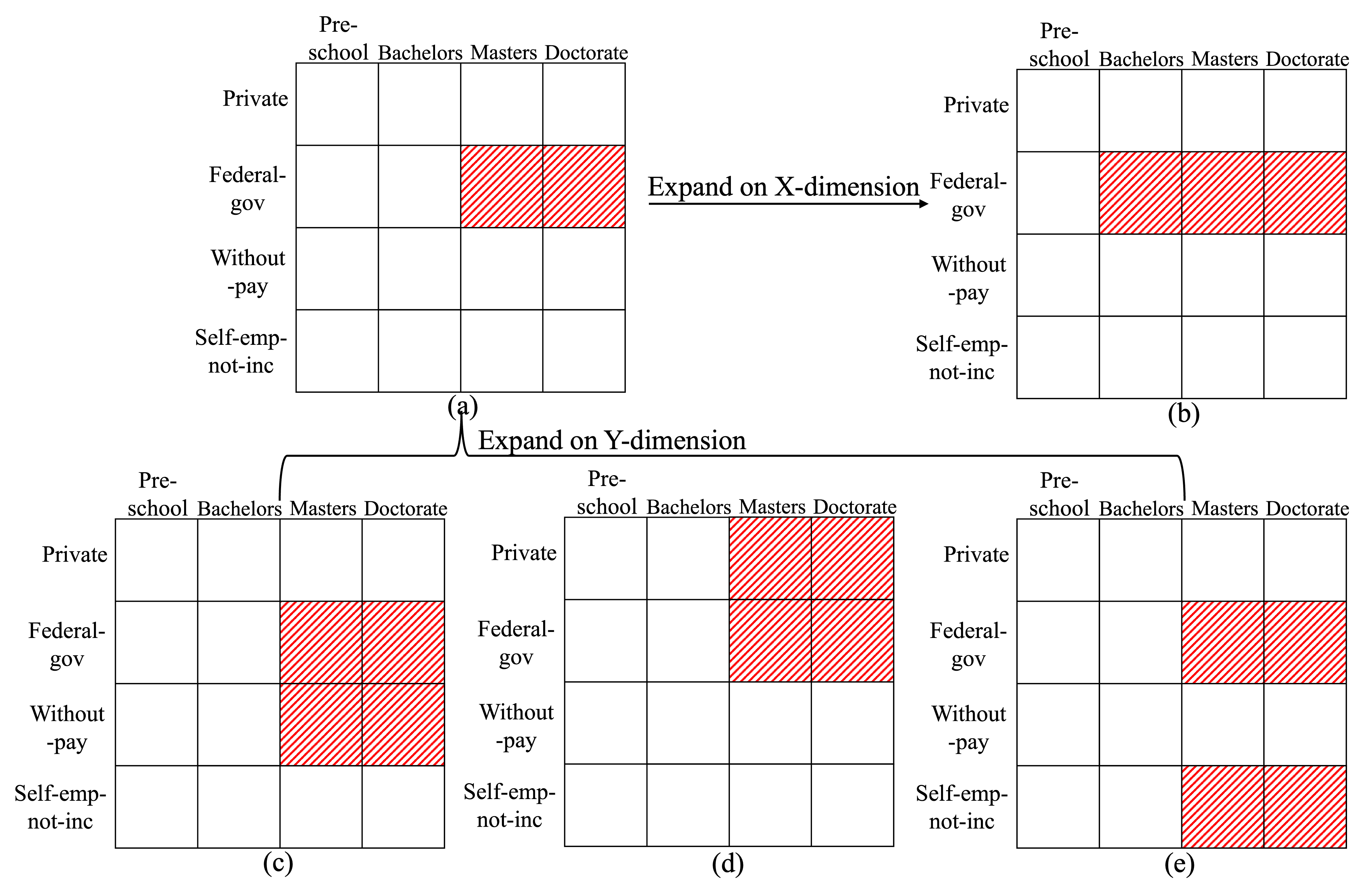}
\caption{A Running Example to Illustrate the Expansion Method}
\label{fig:expansion_running_example}
\end{figure}

We need to tackle two challenges in this method, how to find the potential nearest neighbor feature value on each dimension and how to determine the best neighbor feature value among all possible choices. 

To illustrate our ideas to tackle those challenges, consider a running example in Figure~\ref{fig:expansion_running_example}(a), where the published intent is denoted by the red region. Here, $\mathcal{D} = D_1 \times D_2$, where $D_1$ (horizontal) is the education level and $D_2$ (vertical) the workclass. Let us first look at dimension $D_1$, where the feature values in the published intent are $U_1=$ \{``Masters'', ``Doctorate''\}. The possible feature values to expand $U_1$ include ``Pre-school'' and ``Bachelors.'' For each of those two, we compute the accumulated distance between the feature value and all feature values in $U_1$. The accumulated distance for a feature value $u \in D_i \setminus U_i$ is defined as
$
accu\_dist(u) = \sum_{u' \in U_i} dist(u, u')
$.
Suppose we assign the indices 1, 2, 3, and 4 to ``Pre-school,'' ``Bachelors,'' ``Masters,'' and ``Doctorate,'' respectively. The accumulated distance between ``Pre-school'' and all the features in $U_1$ is $5$.
Similarly, the accumulated distance for ``Bachelors'' with respect to $U_1$ is 3. The feature value with the shortest accumulated distance is selected as the nearest neighbor feature value on $D_1$, which is ``Bachelors.'' 
On dimension $D_2$, there is no semantic distance among feature values. In other words, every pair of feature values on $D_2$ has the same distance. Thus, regarding the features in the published intent on dimension $D_2$, which is $U_2= $ \{``Federal-gov''\}, all the remaining features ``Without-pay,'' ``Private,'' and ``Self-emp-not-inc'' are treated as potential expanded feature values. 

After iterating through every dimension, we have four neighbor feature values for possible expansion: ``Bachelors'' on $D_1$ (the resulting published intent is indicated in Figure~\ref{fig:expansion_running_example}(b)), ``Without-pay'' (Figure~\ref{fig:expansion_running_example}(c)), ``Private'' (Figure~\ref{fig:expansion_running_example}(d)), and ``Self-emp-not-inc'' (Figure~\ref{fig:expansion_running_example}(e)) on $D_2$. 

To determine the best feature value, we develop a scoring function to comprehensively measure the effectiveness of adding the feature value to the published intent. Given a potential feature value $u_{ij}$ to be added, which is a feature value on dimension $D_i$, the scoring considers two factors. First, we consider the additional number of records added to the buyer's published intent after adding the feature value. Given feature value $u_{ij}$ and the current published intent $U_1 \wedge \cdots \wedge U_i \wedge \cdots \wedge U_n$, the addition can be expressed as
$
addition(u_{ij}) = |\{x_k\}|_{x_k \in U_1 \wedge \cdots \wedge \{u_{ij}\} \cdots \wedge U_n} 
$.
Second, we also consider the contribution to satisfy the constraints in Equation~\ref{PI-uniform_objective} (PI-uniform attack) or Equation~\ref{EM_objective} (efficiency maximization attack), which is measured as the increase in the right-hand side of the constraint after adding the feature value and can be expressed as follows.  For the PI-uniform attack, 
\begin{equation}
    increase(u_{ij}) = |U_1 \wedge \cdots \wedge U_{i-1} \wedge \{u_{ij}\} \wedge U_{i+1} \cdots \wedge U_n|
\end{equation}
For the efficiency maximization attack,
\begin{equation}
    increase(u_{ij}) = \sum_{t\in U_1 \times \cdots \times U_{i-1} \times \{u_{ij}\} \times U_{i+1} \times \cdots \times U_n}f_\mathcal{D}(t)
    \label{EM_attack_increase}
\end{equation}

For each neighbor feature value, we apply min-max normalization to the associated addition of records and increase in satisfying the constraint. Then, we compute the score of the neighbor feature value $u_{ij}$ by
\begin{equation}\label{eq:expansion_score_equation}
\begin{aligned}
    score(u_{ij}) &= \alpha \cdot \left(\max_{u \in \cup_{i=1}^n (D_i \setminus U_i)}\{{\widehat{addition}(u)}\} -\widehat{addition}(u_{ij})\right) \\
    &\quad+ (1-\alpha) \cdot \widehat{increase}(u_{ij})
\end{aligned}
\end{equation}
where $\alpha$ is the weight assigned to the evaluation factor, $\widehat{addition}(u_{ij})$ and $\widehat{increase}(u_{ij})$ are the normalized addition of records and increase in satisfying the constraint for $u_{ij}$, respectively, $\max_{u \in \cup_{i=1}^n (D_i \setminus U_i)}\{{\widehat{addition}(u)}\}$ is the maximum of the addition of all possible neighbor feature values.

For the efficiency maximization attack, in the worst case, the buyer has to claim the entire dataset as the published intent and the attainable lower bound of the attacker's confidence is
$$
Conf_{\text{PI}, f_\mathcal{D}}^{\text{Eff}}(x) \geq \frac{\max_{r \in V_1 \times \cdots \times V_n}f_{\mathcal{D}}(r)}{\sum_{t\in D_1 \times \cdots \times D_n}f_\mathcal{D}(t)} 
$$

\subsection{Purchased Record Inference Attack}\label{sec:PRI_allocation_methods}

The ideas can be extended to tackle the problem of privacy protection against purchased record inference attacks.

Given the objective in Equation~\ref{PRI_objective}, in order to make the p-value of $f_\mathcal{P}(x)$ greater than or equal to $1-\lambda$ for any $x \in V_1 \times \cdots \times V_n$, we first need to decide the buyer's published intent (the attacker infers it as the pseudo published intent $\widehat{PI}$). Thus, given the buyer's true intent, we first leverage the proposed expansion method to derive the published intent. 

Then, given the total number of records, $q$, the buyer aims to buy and the published intent $PI = U_1 \wedge \ldots \wedge U_n$, how to allocate the $q$ records to every feature value combination in $U_1 \wedge \ldots \wedge U_n$ becomes a challenge. To solve the challenge, we propose four allocation methods:

\textbf{Monte Carlo Simulation (MC Simulation):} Generate $Z$ sets, $\mathcal{X}_1, \mathcal{X}_2, \mathcal{X}_3, ... \mathcal{X}_Z$, with each set containing $q$ records sampled from the published intent. Then, remove sets that do not satisfy the privacy constraint, i.e. the p-value of $f_\mathcal{P}(x)$ less than $1-\lambda$ for any $x \in V_1 \times \cdots \times V_n$.

For each set remaining, we compute the utility, i.e. the ratio of the number of records in the true intent against that in the published intent. Given one of the remaining sets denoted as $\mathcal{X}_i$, we have:
\begin{equation}\label{eq:utility_computation} 
utility(\mathcal{X}_i) = \frac{|\{r\}_{r \in \mathcal{X}_i, r \in V_1 \times \cdots \times V_n}|}{|\mathcal{X}_i|}
\end{equation}

Considering utility as a primary consideration, where the buyer seeks to maximize the ratio of the number of records in the true intent against that in the published intent, the set with the highest utility is selected as the purchased set of records.

\textbf{Markov Chain Monte Carlo (MCMC):} We first initialize the purchased set $\mathcal{X} = \{x_1, x_2, ..., x_q\}$ by sampling $q$ records based on data distribution $f_{\mathcal D, \widehat{PI}}$ as indicated in Equation~\ref{eq:ground_truth_data_distribution_PI}.


Then, we define a set of possible moves. Given the initialized set of record $\mathcal{X}$ containing the records that the data buyer really wants (records in the buyer's true intent) and some disguising records that the data buyer uses to hide the privacy, all the possible moves can be summarized into 4 groups: 1) remove a disguising record and add a record in the buyer's true intent, 2) remove a disguising record and add a different disguising record, 3) remove a record in the buyer's true intent and add a disguising record, 4) remove a record in the buyer's true intent and add a different record in the buyer's true intent. Note that multiple move choices exist under each group by selecting different disguising records or records in the buyer's true intent.

For each iteration, we randomly select one move and change the current $\mathcal{X}$ to $\hat{\mathcal{X}}$ based on the move. We first check whether $\hat{\mathcal{X}}$ satisfies the privacy constraint. If the privacy constraint is satisfied, we compute the acceptance probability of $\hat{\mathcal{X}}$ as follows:
$$
accept\_prob = \min(1,\frac{utility(\hat{\mathcal{X})}}{utility(\mathcal{X})})
$$
where the utility function is elaborated in Equation~\ref{eq:utility_computation}.

We accept $\hat{\mathcal{X}}$ with the computed acceptance probability, i.e. there is a $accept\_prob$ probability of transitioning from $\mathcal{X}$ to $\hat{\mathcal{X}}$.

We continue selecting the next move and updating the purchased set of records until the privacy constraint is no longer satisfied, i.e. the p-value of $f_\mathcal{P}(x)$ less than $1-\lambda$ for any $x \in V_1 \times \cdots \times V_n$. Note that during implementation, the MCMC Algorithm terminates when the minimum difference between the privacy threshold $\lambda$ and the current confidence of an attacker on a record $x \in V_1 \times \cdots \times V_n$ is less than a threshold $\epsilon$.

\textbf{Greedy Markov Chain Monte Carlo (G-MCMC):} The only difference between this approach and normal Markov Chain Monte Carlo is that when we define the set of possible moves, we only define the most greedy moves that improve the utility, i.e. remove a disguising record and add a record in the buyer's true intent.

\textbf{Genetic Sampling:} Initialize $T$ parent sets $\mathcal{X}_1, \mathcal{X}_2, ..., \mathcal{X}_T$ where each set contains $q$ records sampled from the published intent. Then, remove sets that do not satisfy the privacy constraint, i.e. the p-value of $f_\mathcal{P}(x)$ less than $1-\lambda$ for any $x \in V_1 \times \cdots \times V_n$. For the remaining sets, we compute the utility based on the utility expression in Equation~\ref{eq:utility_computation}.

Then, we select top $R$ sets in terms of utility to form a group. For every pair (parent A, B) of sets in the group, we leverage crossover to generate two children sets (children A, B). We add all the generated children to a new population pool.

We repeat the above steps on the new population pool for a specified number of iterations $W$.

Finally, among all the sets satisfying the privacy constraint in the ultimate population pool, we select the set with the highest utility to be the buyer's purchased set.

\section{Experimental Results}\label{sec:exp}
We conduct extensive experiments using both real and synthetic datasets to investigate the following questions.

\textbf{RQ1:} To what extent does the proposed expansion method effectively generate privacy-preserving and high-utility published intent?

\textbf{RQ2:} To what extent does each allocation method effectively generate privacy-preserving and high-utility set of data records for purchase by the buyer?

\textbf{RQ3:} What is the effect of data space dimensionality on the published intent in privacy and utility?

\textbf{RQ4:} What is the effect of buyer's true intent on the published intent produced by the expansion method?

\textbf{RQ5:} What are the effects of the privacy threshold $\lambda$ and the weight parameter $\alpha$ in the expansion method?

\subsection{Experimental Setup}

The \textbf{Adult Dataset~\cite{misc_adult_2}} is adopted from the UC Irvine machine learning repository~\cite{asuncion2007uci} derived from US census data. Our analysis focuses on five attributes: age, ethnicity, gender, hours-per-week, and income. The age attribute is transformed into an ordinal scale: \{``childhood'' (0-17], ``young adult'' (17-24], ``working adult'' (24-61], ``retirement'' $>$61\}. Similarly, attribute hours-per-week is discretized into: \{``part-time'' [0-34], ``full-time'' (34-40], ``overtime'' $>$40\}. After removing the records with missing values, the dataset comprises 30,162 valid records. 
As the cost of each record is unspecified, a uniform cost of 1 is assigned to each record.

We generated a \textbf{synthetic dataset} mimicking the structure of the Adult Dataset, sharing the five attributes. The number of distinct feature values per dimension matches the Adult Dataset. The same data processing technique is applied. Regarding the data distribution, the record frequencies are sampled from a Gaussian distribution (mean=1000, std=300). 
The associated record costs are sampled from another Gaussian distribution (mean=20, std=5, lower bound=1).

\begin{table*}[ht]
  \centering
  \caption{Effectiveness of Expansion Method on PI-uniform Attack and Efficiency Maximization Attack on the Adult Dataset. ``PI-uniform'' refers to PI-uniform Attack, and ``EM'' stands for Efficiency Maximization Attack.  Variants of EM-(fc, f, c) represent the attacker's background knowledge: `fc' includes data distribution and cost, `f' is only distribution, and `c' is only cost. PI and TI denote Published Intent and True Intent, respectively.}
  \label{tab:RQ1_table_expansion_effectiveness_adult_dataset}
  \begin{tabularx}{\textwidth}{XXXXXXXXXXX}
       \toprule
    \textbf{Setting} & 
    \textbf{Attack Type} & \textbf{Prot. Method} & \textbf{Conf. LB} & \textbf{Conf. UB} & \textbf{Total Cost} & \textbf{Cost (TI)} & \textbf{Cost Ratio (TI)} & \textbf{\# Records (PI)} & \textbf{\# Records (TI)} & \textbf{Utility (TI/PI)}  \\
    \midrule
   \multirow{16}{1.5cm}{Adult Dataset, Unit Cost of 1, True Intent Size is 1} & \multirow{4}{2cm}{PI-uniform} & w/o protection & 100\% & 100\% & 56.0 & 56.0 & 100\% & 56 & 56 & 100\% \\ \cmidrule(lr){3-11}
    & & Expansion  & 25\% & 25\% & 58.0 & 56.0 & 96.6\% & 58 & 56  & 96.6\% \\  \cmidrule(lr){2-11}
    & \multirow{4}{2cm}{EM-fc} & w/o protection & - & 100\%& 56.0& 56.0& 100\%& 56& 56& 100\%\\ \cmidrule(lr){3-11}
    & & Expansion & - & 10.1\% & 557.0&  56.0& 10.1\%& 557& 56& 10.1\%\\ \cmidrule(lr){2-11}
    & \multirow{4}{2cm}{EM-f} & w/o protection& - & 100\%& 56.0& 56.0& 100\%& 56& 56& 100\%\\ \cmidrule(lr){3-11}
    & & Expansion & - & 10.1\% & 557.0&  56.0& 10.1\%& 557& 56& 10.1\%\\
    \cmidrule(lr){2-11}
    &  \multirow{4}{2cm}{EM-c} & w/o protection & - & 100\%& 56.0& 56.0& 100\%& 56& 56& 100\%\\ \cmidrule(lr){3-11}
    & & Expansion & - & 25\% & 58.0&56.0 &96.6\% &58 &56 & 96.6\%\\
    \midrule
       \multirow{16}{1.5cm}{Adult Dataset, Unit Cost of 1, True Intent Size is 2} & \multirow{4}{2cm}{PI-uniform} & w/o protection & 50\% & 100\% & 83.0 & 83.0 & 100\% & 83 & 83 & 100\% \\ \cmidrule(lr){3-11}
    & & Expansion  & 12.5\% & 25\% & 85.0 & 83.0 & 97.6\% & 85 & 83  & 97.6\% \\  \cmidrule(lr){2-11}
    & \multirow{4}{2cm}{EM-fc} & w/o protection & - & 67.5\% & 83.0 & 83.0 & 100\% & 83 & 83 & 100\% \\ \cmidrule(lr){3-11}
    & & Expansion & - & 8.4\% & 669.0 & 83.0 & 12.4\% & 669 & 83 & 12.4\% \\ \cmidrule(lr){2-11}
    & \multirow{4}{2cm}{EM-f} & w/o protection & - & 67.5\% & 83.0 & 83.0 & 100\% & 83 & 83 & 100\% \\ \cmidrule(lr){3-11}
    & & Expansion & - & 8.4\% & 669.0 & 83.0 & 12.4\% & 669 & 83 & 12.4\% \\
    \cmidrule(lr){2-11}
    &  \multirow{4}{2cm}{EM-c} & w/o protection & - & 50\% & 83.0 & 83.0 & 100\% & 83 & 83 & 100\% \\ \cmidrule(lr){3-11}
    & & Expansion & - & 25\% & 84.0 & 83.0 & 98.8\% & 84 & 83 & 98.8\% \\
    \bottomrule
  \end{tabularx}
\end{table*}

As case studies, we define two distinct buyer true intents: [(``working adult'', ``Black'', ``Female'', ``full-time'', ``$>$50K'')] with size 1 and [(``working adult'', ``Black'', ``Female'', ``full-time'', ``$>$50K''), (``working adult'', ``Asian-Pac-Islander'', ``Female'', ``full-time'', ``$>$50K'')] with size 2. We set $\lambda$ to 30\%. To maximize the utility of the published intents, i.e. the ratio of the number of records in the true intent against that in the published intent, for the Adult Dataset with a true intent size of 1, we set $\alpha$ to 0.5 for PI-uniform attack, EM attack-fc, EM attack-f, and EM attack-c, where EM attack-fc, EM attack-f, and EM attack-c refer to efficiency maximization attacks given the attacker background knowledge about both the data distribution and cost, only the data distribution, and only the cost. For the Adult Dataset with a true intent size of 2, we set $\alpha$ to 1.0, 0.6, 0.6, and 0.5 for PI-uniform attack, EM attack-fc, EM attack-f, and EM attack-c, respectively. For the synthetic dataset with a true intent size of 1, we set $\alpha$ to 0.8, 0.4, 0.4, and 0.8 for PI-uniform attack, EM attack-fc, EM attack-f, and EM attack-c, respectively. For the synthetic dataset with a true intent size of 2, we set $\alpha$ to 0.6, 0.5, 0.4, and 0.7 for PI-uniform attack, EM attack-fc, EM attack-f, and EM attack-c, respectively.

For purchased record inference attack, $\mathcal{L}$, which denotes the number of sets used for Monte Carlo simulation and permutation testing, is set to 100,000.

For MC Simulation, the total number of sets $Z$ is set to 100,000. For MCMC, the minimum difference between the privacy threshold and the current confidence of the attacker $\epsilon$ is set to 0.001 for all case studies, while for G-MCMC, $\epsilon$ is set to 0.07, 0.01, 0.005, and 0.01 for the Adult Dataset with a true intent size of 1, Adult Dataset with a true intent size of 2, synthetic dataset with a true intent size of 1, and synthetic dataset with a true intent size of 2, respectively. For genetic sampling, the number of parents $T$ is set to 50, the number of top utility sets selected for crossover $R$ is set to 10, and the number of iterations $W$ is set to 30. 






\begin{table*}[ht]
  \centering
  \caption{Effectiveness of Expansion Method on PI-uniform Attack and Efficiency Maximization Attack on the synthetic dataset. ``PI-uniform'' refers to PI-uniform Attack, and ``EM'' stands for Efficiency Maximization Attack. Variants of EM-(fc, f, c) represent the attacker's background knowledge: `fc' includes data distribution and cost, `f' is only distribution, and `c' is solely cost. PI and TI denote Published Intent and True Intent, respectively.}
  \label{tab:RQ1_table_expansion_effectiveness_synthetic_data}
  \begin{tabularx}{\textwidth}{XXXXXXXXXXX}
    \toprule
    \textbf{Setting} & 
    \textbf{Attack Type} & \textbf{Prot. Method} & \textbf{Conf. LB} & \textbf{Conf. UB} & \textbf{Total Cost} & \textbf{Cost (TI)} & \textbf{Cost Ratio (TI)} & \textbf{\# Records (PI)} & \textbf{\# Records (TI)} & \textbf{Utility (TI/PI)}  \\
    \midrule
       \multirow{16}{1.5cm}{Synthetic Dataset, Gaussian Cost, True Intent Size is 1} & \multirow{4}{2cm}{PI-uniform} & w/o protection & 100\% & 100\% & 31773.0 & 31773.0 & 100\% & 1513 & 1513 & 100\% \\ \cmidrule(lr){3-11}
    & & Expansion  & 25\% & 25\% & 63430.0 & 31773.0 & 50.1\% & 3222 & 1513  & 47.0\% \\  \cmidrule(lr){2-11}
    & \multirow{4}{2cm}{EM-fc} & w/o protection & - & 100\% & 31773.0 & 31773.0 & 100\% & 1513 & 1513 & 100\% \\ \cmidrule(lr){3-11}
    & & Expansion & - & 29.5\% & 107825.0 & 31773.0 & 29.5\% & 5715 & 1513 & 26.5\% \\ \cmidrule(lr){2-11}
    & \multirow{4}{2cm}{EM-f} & w/o protection & - & 100\% & 31773.0 & 31773.0 & 100\% & 1513 & 1513 & 100\% \\ \cmidrule(lr){3-11}
    & & Expansion & - & 26.5\% & 107825.0 & 31773.0 & 29.5\% & 5715 & 1513 & 26.5\% \\
    \cmidrule(lr){2-11}
    &  \multirow{4}{2cm}{EM-c} & w/o protection & - & 100\% & 31773.0 & 31773.0 & 100\% & 1513 & 1513 & 100\% \\ \cmidrule(lr){3-11}
    & & Expansion & - & 28.8\% & 63430.0 & 31773.0 & 50.1\% & 3222 & 1513 & 47.0\% \\
    \midrule
       \multirow{16}{1.5cm}{Synthetic Dataset, Gaussian Cost, True Intent Size is 2} & \multirow{4}{2cm}{PI-uniform} & w/o protection & 50\% & 100\% & 58053.0 & 58053.0 & 100\% & 2973 & 2973 & 100\% \\ \cmidrule(lr){3-11}
    & & Expansion  & 12.5\% & 25\% & 154470.0 & 58053.0 & 37.6\% & 8153 & 2973  & 36.5\% \\  \cmidrule(lr){2-11}
    & \multirow{4}{2cm}{EM-fc} & w/o protection & - & 54.7\% & 58053.0 & 58053.0 & 100\% & 2973 & 2973 & 100\% \\ \cmidrule(lr){3-11}
    & & Expansion & - & 25.2\% & 125874.0 & 58053.0 & 46.1\% & 6380 & 2973 & 46.6\% \\ \cmidrule(lr){2-11}
    & \multirow{4}{2cm}{EM-f} & w/o protection & - & 50.9\% & 58053.0 & 58053.0 & 100\% & 2973 & 2973 & 100\% \\ \cmidrule(lr){3-11}
    & & Expansion & - & 29.2\% & 98329.0 & 58053.0 & 59.0\% & 5179 & 2973 & 57.4\% \\
    \cmidrule(lr){2-11}
    &  \multirow{4}{2cm}{EM-c} & w/o protection & - & 53.8\% & 58053.0 & 58053.0 & 100\% & 2973 & 2973 & 100\% \\ \cmidrule(lr){3-11}
    & & Expansion & - & 29.6\% & 75150.0 & 58053.0 & 77.2\% & 3954 & 2973 & 75.2\% \\
    \bottomrule
  \end{tabularx}
\end{table*}

\subsection{Privacy and Utility of Published Intent}\label{sec:RQ1}

The experimental results are shown in Table~\ref{tab:RQ1_table_expansion_effectiveness_adult_dataset} for the Adult Dataset and Table~\ref{tab:RQ1_table_expansion_effectiveness_synthetic_data} for the synthetic dataset. 

In PI-uniform attacks, the absence of protection results in the attacker's upper bound confidence reaching 100\% for both datasets and for both true intent sizes of 1 and 2. This underscores the crucial role of the expansion method in safeguarding privacy. 
Employing the expansion method to counter PI-uniform attacks leads to significant reduction of the confidence upper bound. At the same time, this privacy protection strategy comes with a tradeoff – it requires the acquisition of additional disguise data, resulting in a slight decrease in utility. On the Adult Dataset, the utility decreases slightly. 

For the synthetic dataset, the utility decreases by 53\% when the true intent size is 1 and by 63.5\% when the true intent size is 2. Notably, the more pronounced reduction in the utility of published intents in the synthetic dataset, compared to the Adult Dataset, can be attributed to the presence of ``empty'' feature value combinations. For example, in the Adult Dataset, there are 0 records falling into the group of (``childhood'', ``white'', ``female'', ``overtime'', ``$>$50k''). In a PI-uniform attack, where an attacker lacks of the knowledge of the ground truth data distribution. Such ``empty'' groups in the published intent enhance privacy without incurring additional disguise records. Such ``empty'' groups are rare in the synthetic dataset, resulting in a greater drop in the utility of published intent. In efficiency maximization attacks, the ``empty'' groups also result in significantly higher utility of the published intent after the expansion method, given the attacker's background knowledge about costs alone. This utility exceeds the utility of the published intent after the expansion method when the attacker has background knowledge about both data distribution and costs, or only data distribution, by over 86\% for the Adult Dataset and over 20\% for the synthetic dataset.

In efficiency maximization attacks, given the attacker's background knowledge about both the data distribution and costs, without protection an attacker's confidence upper bound reaches 100\% when the true intent size is 1 and 67.5\% when the true intent size is 2 for the Adult Dataset. Regarding the synthetic dataset, without protection an attacker's confidence upper bound reaches 100\% when the true intent size is 1 and 54.7\% when the true intent size is 2. The difference in confidence upper bound between true intent size of 1 and true intent size of 2 arises from the inherent characteristics of the efficiency maximization attack. The attack confidence is calculated based on the maximum $f_{\mathcal{D}}(r)$ for record $r$ in the buyer's true intent, divided by the sum of $f_{\mathcal{D}}(t)$ for every record $t$ in the published intent, as indicated in Equation~\ref{EM_objective}. Without protection and with a true intent size of 1, the confidence is necessarily 100\% since the published intent contains only this single true intent. However, with a true intent size of 2 and no protection, the numerator is the maximum between the two true intents, while the denominator is the sum of the two, resulting in a lower attack confidence. In essence, the presence of multiple true intents balances each other, leading to a lower attacker confidence even without protection. 

For the Adult Dataset, upon implementing the expansion method, the attack confidence decreases by 89.9\% when the true intent size is 1 and by 59.1\% when the true intent size is 2. The increase in privacy comes at the cost of utility, i.e. a decrease by 89.9\% and 87.6\%, respectively. Regarding the synthetic dataset, the attack confidence decreases by 70.5\% when the true intent size is 1 and by 29.5\% when the true intent size is 2. 
The increase in privacy comes at the cost of utility, i.e. a decrease by 73.5\% and 53.4\%, respectively.

\subsection{Effectiveness Comparison of Different Allocation Methods}
\begin{table*}[ht]
  \centering
  \caption{Effectiveness Comparison of Different Allocation Methods for Purchased Record Inference Attack. For each method, the mean and standard deviation of ten runs are reported. The best performance in the privacy, utility, and runtime columns is highlighted in boldface.}
  \label{tab:RQ2_different_allocation_methods_PRI}
  \begin{tabularx}{\textwidth}{XXXXXXX}
    \toprule
    \textbf{Setting} & 
     \textbf{Protection Method}  & \textbf{Confidence Upper Bound} & \textbf{\# Records (PI)} & \textbf{\# Records (TI)} & \textbf{Utility (TI/PI)} & \textbf{Time} \\
    \midrule
   \multirow{8}{1.5cm}{Adult Dataset, True Intent Size is 1} & No Protection & 100.0\% & 61.0 & 56.0 & 91.8\% & -\\ \cmidrule(lr){2-7}
   & MC Simulation& 23.6\%$\pm$0.0\%  & 61.0$\pm$0.0 & 5.0$\pm$0.0 & \textbf{8.2\%$\pm$0.0\%} & 24.154$\pm$0.144\\  \cmidrule(lr){2-7}
   & MCMC & \textbf{6.4\%$\pm$6.8\%}  & 61.0$\pm$0.0 & 3.1$\pm$0.9 & 5.1\%$\pm$1.5\% & 0.012$\pm$0.001\\ \cmidrule(lr){2-7}
   & G-MCMC & 23.6\%$\pm$0.0\%  & 61.0$\pm$0.0 & 5.0$\pm$0.0 & \textbf{8.2\%$\pm$0.0\%} & \textbf{0.001$\pm$0.000}\\ \cmidrule(lr){2-7}
   & Genetic Sampling& 23.6\%$\pm$0.0\%  & 61.0$\pm$0.0 & 5.0$\pm$0.0 & \textbf{8.2\%$\pm$0.0\%} & 0.301$\pm$0.003\\ 
    \midrule
       \multirow{8}{1.5cm}{Adult Dataset, True Intent Size is 2} & No Protection & 100.0\% &  120.0 & 83.0 & 69.2\% & -\\ \cmidrule(lr){2-7}
   & MC Simulation& 29.3\%$\pm$0.0\%  & 120.0$\pm$0.0 & 13.0$\pm$0.0 & \textbf{10.8\%$\pm$0.0\%} & 47.521$\pm$0.290\\  \cmidrule(lr){2-7}
   & MCMC & \textbf{2.7\%$\pm$4.2\%}  & 120.0$\pm$0.0 & 4.0$\pm$2.9 & 3.3\%$\pm$2.4\% & 0.088$\pm$0.007\\ \cmidrule(lr){2-7}
   & G-MCMC& 29.3\%$\pm$0.0\%  & 120.0$\pm$0.0 & 12.7$\pm$0.5 & 10.6\%$\pm$0.4\% & \textbf{0.004$\pm$0.002}\\ \cmidrule(lr){2-7}
   & Genetic Sampling& 28.9\%$\pm$1.1\%  & 120.0$\pm$0.0 & 12.9$\pm$0.3 & \textbf{10.8\%$\pm$0.2\%} & 1.118$\pm$0.010\\ 
    \midrule
       \multirow{8}{1.5cm}{Synthetic Dataset, True Intent Size is 1} & No Protection & 100.0\% &  1516.0 & 1513.0& 99.8\% & -\\ \cmidrule(lr){2-7}
   & MC Simulation & 29.6\%$\pm$0.0\%  & 1516.0$\pm$0.0 & 394.0$\pm$0.0 & \textbf{26.0\%$\pm$0.0\%} & 333.333$\pm$13.589\\  \cmidrule(lr){2-7}
   & MCMC & \textbf{11.6\%$\pm$11.2\%}  & 1516.0$\pm$0.0 & 375.3$\pm$18.6 & 24.8\%$\pm$1.2\% & 0.565$\pm$0.112\\ \cmidrule(lr){2-7}
   & G-MCMC & 29.6\%$\pm$0.0\%  & 1516.0$\pm$0.0 & 394.0$\pm$0.0 & \textbf{26.0\%$\pm$0.0\%} & \textbf{0.009$\pm$0.002}\\ \cmidrule(lr){2-7}
   & Genetic Sampling& 29.1\%$\pm$0.9\%  & 1516.0$\pm$0.0 & 393.8$\pm$0.4 & \textbf{26.0\%$\pm$0.0\%} & 6.300$\pm$0.095\\ 
    \midrule
       \multirow{8}{1.5cm}{Synthetic Dataset, True Intent Size is 2} & No Protection & 100.0\% & 2979.0 & 2973.0 & 99.8\% & -\\ \cmidrule(lr){2-7}
   & MC Simulation& 29.2\%$\pm$0.0\%  & 2979.0$\pm$0.0 & 918.0$\pm$0.0 & \textbf{30.8\%$\pm$0.0\%} & 640.229$\pm$2.150\\  \cmidrule(lr){2-7}
   & MCMC& \textbf{7.7\%$\pm$9.9\%}  & 2979.0$\pm$0.0 & 853.7$\pm$32.4 & 28.7\%$\pm$1.1\% & 1.735$\pm$0.034\\ \cmidrule(lr){2-7}
   & G-MCMC& 29.2\%$\pm$0.0\%  & 2979.0$\pm$0.0 & 916.5$\pm$2.6 &  \textbf{30.8\%$\pm$0.1\%} & \textbf{0.114$\pm$0.099}\\ \cmidrule(lr){2-7}
   & Genetic Sampling& 26.4\%$\pm$1.5\%  & 2979.0$\pm$0.0 & 911.0
$\pm$7.2 & 30.6\%$\pm$0.2\% & 11.566$\pm$3.704\\
    \bottomrule
\end{tabularx}
\end{table*}

Concerning purchased record inference attacks, we assess the effectiveness of four proposed allocation methods in generating privacy-preserving and high-utility set of records for purchase by the data buyer. The experimental results are shown in Table~\ref{tab:RQ2_different_allocation_methods_PRI}.

In purchased record inference attacks, the absence of protection results in the attacker's upper bound confidence reaching 100\% for both datasets. This underscores the crucial role of the allocation method in safeguarding privacy. 

\begin{table*}[t]
  \centering
\caption{Effect of Dimensionality. Change in Lower/Upper Bound Is the Difference Between Projected Attacker’s Confidence and the Attacker’s Confidence Before Projection.}
  \label{tab:RQ3_dimension_removal}
  \begin{tabularx}{\textwidth}{XXXXXXXXXXXXX}
    \toprule
    \textbf{Attack Type} & \textbf{Dim. Reduced} & \textbf{Proj. Conf. LB} & \textbf{Change in LB} & \textbf{Proj. Conf. UB} & \textbf{Change in UB} & \textbf{Proj. \# Records (PI)} & \textbf{Updated Proj. \# Records (PI)} & \textbf{Proj. \# Records (TI)} & \textbf{Proj. Utility (TI/PI)}  & \textbf{Updated Proj. Utility (TI/PI)} & \textbf{Proj. PI Size} & \textbf{Updated Proj. PI Size}\\
    \midrule
    \multirow{9}{1.3cm}{PI-uniform Attack} & Age & 50\% & 37.5\% & 100\% & 75\% & 85 & 101 & 85 & 100\% & 84.2\%& 2 & 8\\ \cmidrule(lr){2-13}
    & Ethnicity & 25\% & 12.5\%& 25\% & 0\% & 85 & 85 & 83 & 97.6\% & 97.6\%& 4 & 4\\ \cmidrule(lr){2-13}
    & Gender & 12.5\% & 0\%& 25\% & 0\% & 85 & 85 & 83 & 97.6\% & 97.6\%& 8 & 8\\ \cmidrule(lr){2-13}
    & Hours-per-week & 12.5\% & 0\%& 25\% & 0\% & 85 & 85 & 83 & 97.6\% & 97.6\%& 8 & 8\\ \cmidrule(lr){2-13}
    & Income & 12.5\% & 0\%& 25\% & 0\% & 85 & 85 & 83 & 97.6\% & 97.6\%& 8 & 8\\
    \midrule
    \multirow{9}{1.3cm}{Efficiency Maximiza -tion Attack} & Age & - & - & 8.7\% & 0.3\% & 669 & 669 & 85 & 12.7\% & 12.7\%& 24 & 24\\ \cmidrule(lr){2-13}
    & Ethnicity & - & - & 13.2\% & 4.8\% & 669 & 390 & 88 & 13.2\% & 22.6\%& 18 & 2\\ \cmidrule(lr){2-13}
    & Gender & - & - & 33.6\% & 25.3\% & 669 & 3226 & 365 & 54.6\% & 11.3\%& 36 & 3\\ \cmidrule(lr){2-13}
    & Hours-per-week & - & - & 12.4\% & 4.0\% & 669 & 669 & 123 & 18.4\% & 18.4\%& 24 & 24\\ \cmidrule(lr){2-13}
    & Income & - & - & 8.4\% & 0\% & 669 & 554 & 83 & 12.4\% & 15.0\%& 72 & 3\\
    \bottomrule
\end{tabularx}
\end{table*}

For the Adult Dataset, MC Simulation, G-MCMC, and Genetic Sampling exhibit comparable results in terms of utility, measured by the ratio of the number of records in the true intent against that in the published intent. MCMC has the worst utility, exhibiting a 3.1\% and 7.5\% decrease for true intent sizes of 1 and 2, respectively. Concerning time, MC Simulation's runtime, requiring 24 seconds for true intent size 1 and 47 seconds for true intent size 2, is notably longer than that of other methods. G-MCMC has the fastest runtime, requiring only 0.001 seconds, while Genetic Sampling requires approximately 1 second. Concerning privacy, MC Simulation, G-MCMC, and Genetic Sampling show an attacker confidence upper bound of 23.6\% and around 29.0\% for true intent sizes of 1 and 2, respectively, all closely approaching the privacy threshold of 30\%. MCMC excels in privacy, exhibiting decreases in attacker confidence of 17.2\% and 26.3\%, respectively.

The synthetic dataset exhibits a similar trend to the Adult Dataset. A slight difference lies in the higher utility of MC Simulation and G-MCMC compared to Genetic Sampling. Another observation is a more apparent gap in runtime between MC Simulation, G-MCMC, and Genetic Sampling, where G-MCMC only requires 0.01 seconds and 0.1 seconds for true intent sizes of 1 and 2, respectively. Genetic Sampling requires 6 seconds and 12 seconds, respectively. MC Simulation performs the worst in runtime, requiring 333 seconds and 640 seconds, respectively.

In summary, our findings indicate that MC Simulation, G-MCMC, and Genetic Sampling consistently achieve comparable results in utility, but G-MCMC consistently outperforms the other two in terms of runtime. Notably, MC Simulation exhibits poor runtime performance. MCMC has the worst performance in utility but is able to generate a more privacy-preserving set of records for purchase by the buyer.

\subsection{Effect of Dimensionality}
We conduct experiments by removing dimensions ``Age,'' ``Ethnicity,'' ``Gender,'' ``Hours-per-week,'' and ``Income,'' respectively, to examine the effect on attacker confidence. The true intent in the reduced space is the projection of the original data space.  That is, with true intent $V_1 \wedge \cdots \wedge V_n$ in the original data space, after dimension $D_i$ is removed, the true intent becomes $V_1 \wedge \cdots V_{i-1} \wedge V_{i+1} \wedge \cdots V_n$. The experiments are conducted on the Adult Dataset with a true intent size of 2. The efficiency maximization attack is conducted given the attacker's background knowledge of both the data distribution and associated costs.

A published intent can also be projected into the subspace reduced, but may no longer meet the privacy threshold. Consequently, we update the projected published intent using our proposed expansion method and compare its utility with that of the initially projected published intent. The results are shown in Table~\ref{tab:RQ3_dimension_removal}.

In the PI-uniform attack, the removal of the ``Age'' dimension has a significant effect, resulting in a substantial increase in the attacker's confidence, both the lower bound (37.5\%) and upper bound (75\%). With the privacy threshold set to 30\%, the projected published intent no longer meets the privacy threshold, necessitating the addition of more disguise records into the published intent. As a consequence, the updated projected number of records in the published intent increases. The updated projected ratio of the number of records in the true intent against that in the published intent decreases by 15.8\%. This observation indicates that, prior to projection, a significant portion of the disguise records used to safeguard the true intent through the expansion method is distributed along the ``Age'' dimension. Thus, the removal of the ``Age'' dimension results in a dramatic increase in attacker confidence.

In the efficiency maximization attack, the removal of the ``Gender'' dimension results in a substantial increase (25.3\%) in attacker confidence, surpassing the 30\% privacy threshold. Consequently, to enhance the protection of the projected published intent, additional disguise records must be incorporated, leading to an increase of the number of records in the published intent. Thus, there is an associated decrease of 43.3\% in the utility.

A noteworthy observation is the comparison between the size of the projected published intent and the size of the updated projected published intent after expansion, specifically after removing the ``Age'' dimension for PI-uniform attack and the ``Gender'' dimension for efficiency maximization attack. In the case of PI-uniform attack, the size increases from 2 to 8, reflecting the expansion method's incorporation of additional feature values into the published intent; thus, the size of the published intent expands. In contrast, for the efficiency maximization attack, the size decreases from 36 to 3. This reduction can be attributed to its inherent characteristics, which necessitates obtaining more records to ensure a sufficiently substantial sum of $f_\mathcal{D}(t)$ for record $t$ in the published intent, as indicated in Equation~\ref{EM_objective}. Consequently, the size of the updated projected published intent may not necessarily increase, given that there are now more records falling within a specific intent due to projection.

\subsection{Effect of True Intent Size}

\begin{figure*}[!htb]
  \centering
  \subfigure[Attacker's Confidence - Age]{\includegraphics[width=0.31\linewidth]{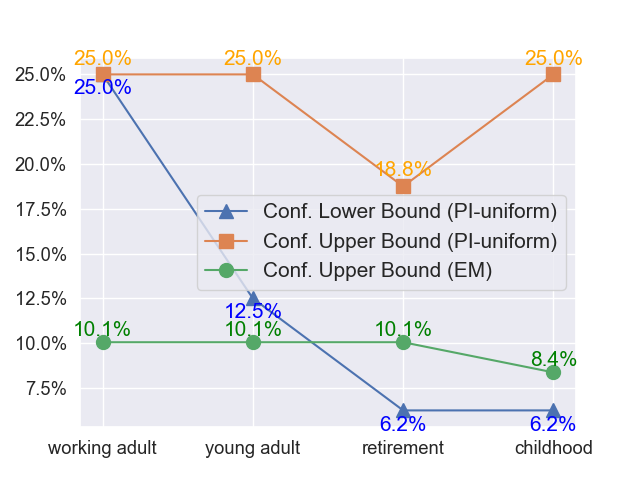}}
  \hfill
  \subfigure[Attacker's Confidence - Ethnicity]{\includegraphics[width=0.31\linewidth]{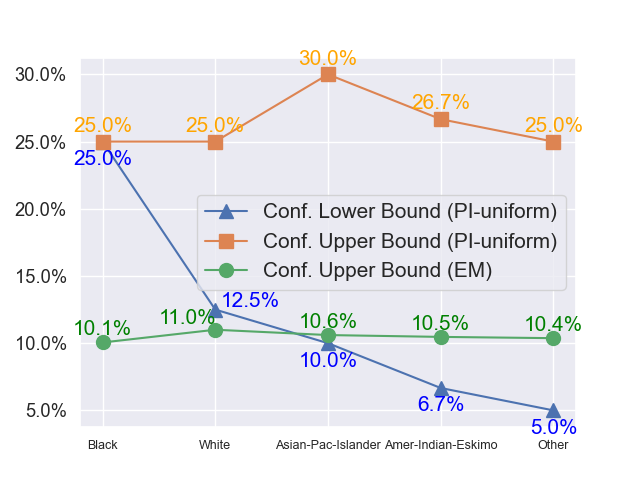}}
  \hfill
  \subfigure[Attacker's Confidence - Hours per Week]{\includegraphics[width=0.31\linewidth]{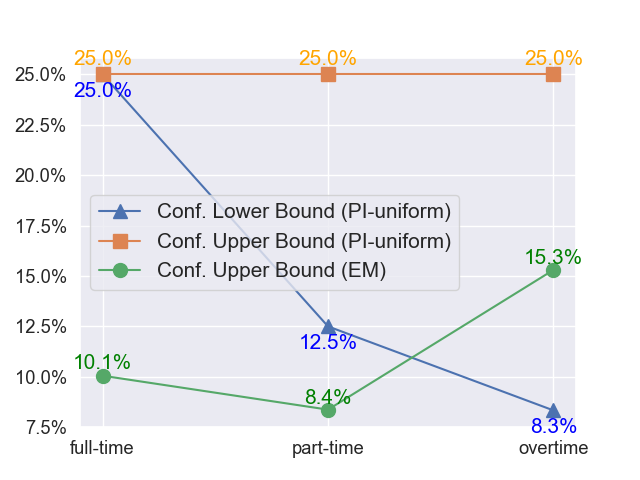}}
  \vspace{0.0cm} 
  \subfigure[Number of Records Purchased - Age]{\includegraphics[width=0.31\linewidth]{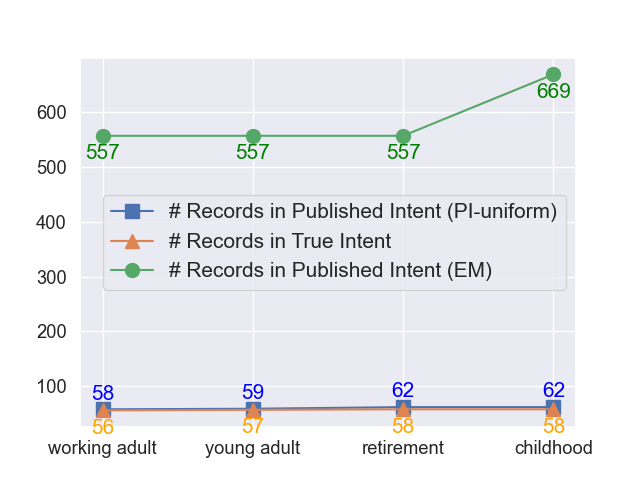}}
  \hfill
  \subfigure[Number of Records Purchased - Ethnicity]{\includegraphics[width=0.31\linewidth]{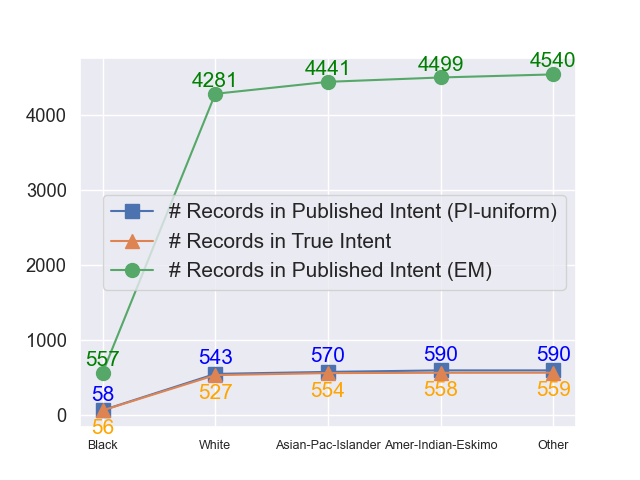}}
  \hfill
  \subfigure[Number of Records Purchased - Hours per Week]{\includegraphics[width=0.31\linewidth]{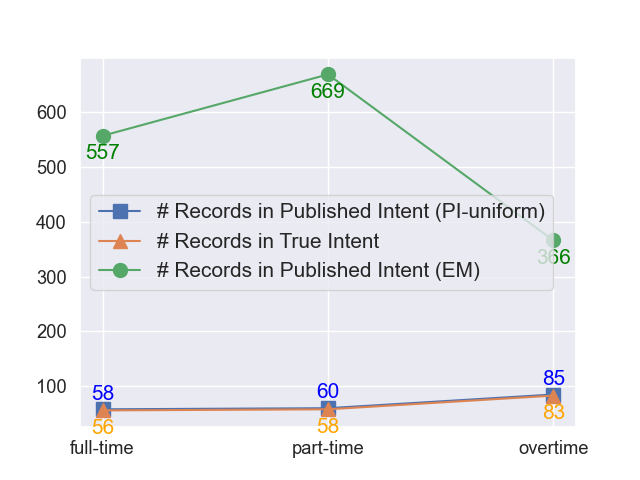}}
  \vspace{0.0cm} 
  \subfigure[Intent Size - Age]{\includegraphics[width=0.31\linewidth]{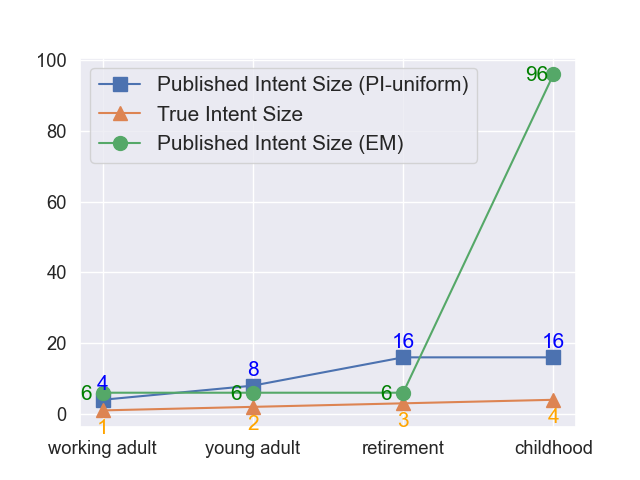}}
  \hfill
  \subfigure[Intent Size - Ethnicity]{\includegraphics[width=0.31\linewidth]{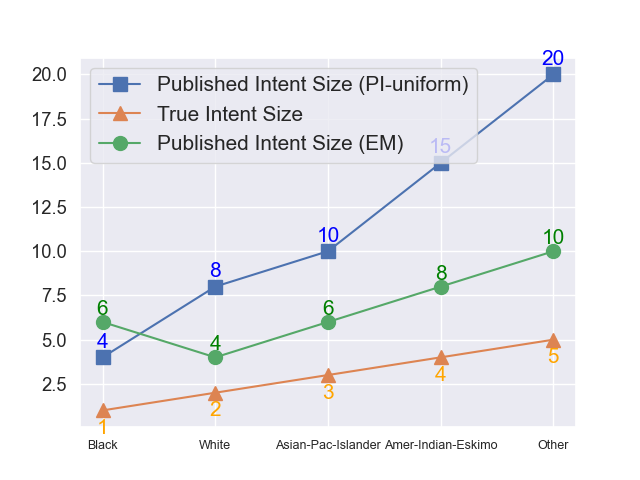}}
  \hfill
  \subfigure[Intent Size - Hours per Week]{\includegraphics[width=0.31\linewidth]{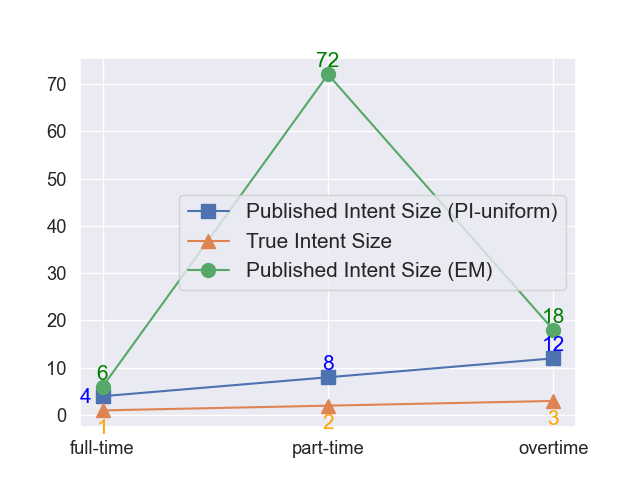}}
\caption{Effect of True Intent Size on Attacker's Confidence, Number of Records Purchased in True Intent (TI) and Published intent (PI), and Published Intent Size regarding Dimensions ``Age'', ``Ethnicity'', and ``Hours per Week'' for PI-uniform Attack (PI-uniform) and Efficiency Maximization Attack (EM).}
  \label{fig:impact-TI-size-attributes}
\end{figure*}

We investigate the effect of buyer's true intent size on the resulting published intent for PI-uniform attack and efficiency maximization attack. For each dimension, we augment the true intent by adding feature values within that dimension. 
The results are shown in Figures~\ref{fig:impact-TI-size-attributes}. In each figure, each feature value on the x-axis denotes that the buyer's true intent on that dimension encompasses the current feature value and all feature values to its left. The experiments are conducted on the Adult Dataset with a true intent size of 1. The efficiency maximization attack is conducted given the attacker's background knowledge of both the data distribution and associated costs.

In the PI-uniform attack, 
we observe that the sizes of the published intent and true intent often consistently increase in the same ratio, with no change in the upper bound confidence. Nevertheless, certain instances deviate from this pattern, where the size of the published intent does not increase despite an increase in the attacker confidence upper bound, which is because the original confidence is excessively low. Moreover, it is evident that the attacker's confidence lower bound decreases as the size of the published intent increases. 

In the efficiency maximization attack, in most instances, the size of the published intent either remains constant or experiences a slight increase with an enlargement of the true intent, all while maintaining a consistent confidence level. 

Notably, when expanding on a new feature value, the new published intent may exhibit higher utility, i.e. lower number of disguise records, even if the original true intent is a proper subset of the expanded true intent. This phenomenon arises due to the greedy nature of the expansion method, 
whereby, at each iteration, the feature value deemed the best may not necessarily be the globally optimal one.

\subsection{Sensitivity Analysis}
Sensitivity analyses are conducted on the Adult Dataset with a true intent size of 2. The efficiency maximization attack is conducted given the attacker's background knowledge of both the data distribution and associated costs.

\begin{figure*}[t]
  \centering
  \begin{minipage}[b]{0.49\textwidth}
    \includegraphics[width=\textwidth]{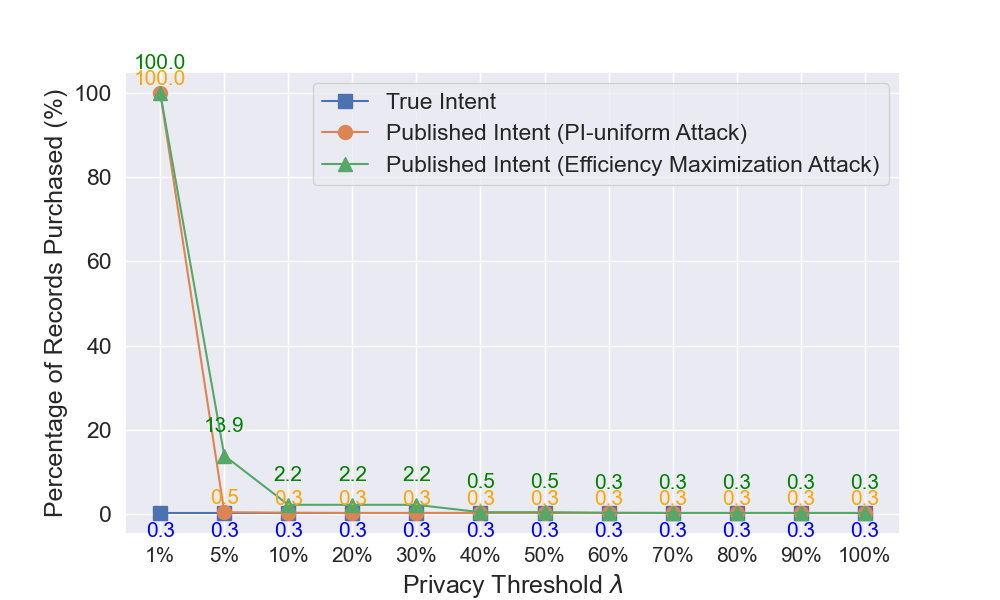}
    \caption{Effect of Privacy Threshold $\lambda$.}
    \label{fig:impact_of_lambda}
  \end{minipage}
  \hfill
  \begin{minipage}[b]{0.49\textwidth}
    \includegraphics[width=\textwidth]{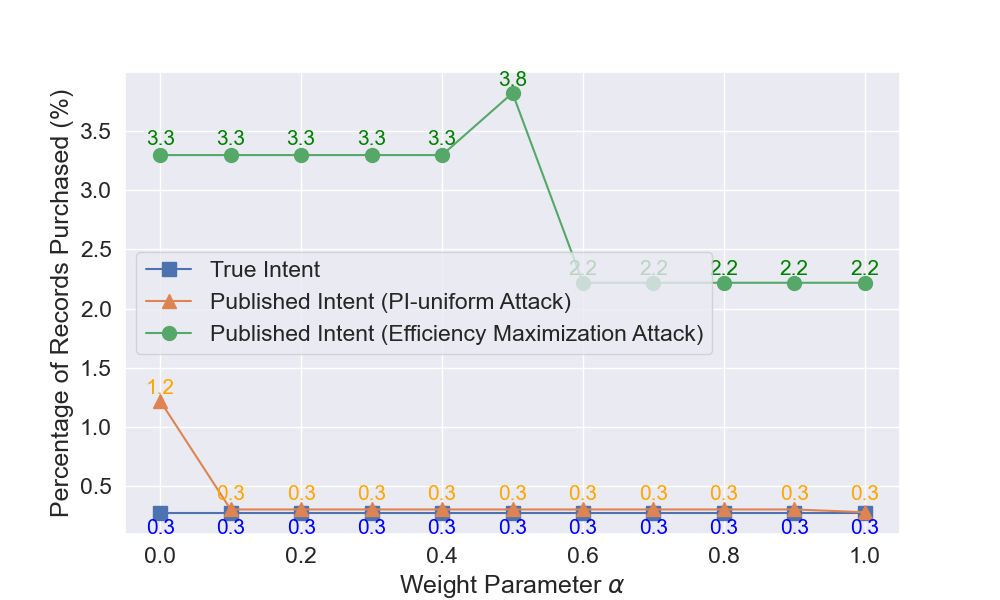}
    \caption{Effect of Weight Parameter $\alpha$}
    \label{fig:impact_of_alpha}
  \end{minipage}
\end{figure*}

\subsubsection{Effect of Privacy Threshold $\lambda$}

The parameter $\lambda$ governs the degree to which the buyer's true intent is protected, with the attacker's confidence constrained to be no greater than $\lambda$. A lower $\lambda$ indicates a higher level of privacy protection. We explore the effect of $\lambda$ on the resulting published intent by varying it from 1\% (the most stringent privacy threshold) to 100\% (indicating no privacy requirement). The results are shown in Figure~\ref{fig:impact_of_lambda}.

In both the PI-uniform attack and the  efficiency maximization attack, adhering to the most rigorous privacy threshold of 1\%, the buyer is compelled to acquire the entire dataset to ensure privacy protection. As the privacy requirement becomes less stringent, the demand for concealing records diminishes quickly. Ultimately, in scenarios where privacy protection is not a concern (when $\lambda$ is set to 100\%), the buyer only needs to buy the data pertaining to the true intent.  The results indicate that privacy preservation for data buyer is highly feasible and economically efficient as long as the privacy requirement is not extremely high.

\subsubsection{Effect of Weight Parameter $\alpha$}

In the expansion method, the parameter $\alpha$ governs the allocation of weights to the total number of records in the published intent (weighted by $\alpha$) and the gain in satisfying the privacy constraint when expanding on the current feature (weighted by $1-\alpha$). We investigate the effect of $\alpha$ on the resulting published intent. The results are shown in Figure~\ref{fig:impact_of_alpha}. 

For the PI-uniform attack with a 30\% privacy threshold, setting $\alpha$ to 1 yields the fewest disguising records purchased. This minimal number of disguising records corresponds to the largest proportion of records indeed belong to the true intent and thus the highest utility. For efficiency maximization attack, with a 30\% privacy threshold, $\alpha$ values greater than or equal to 0.6 result in the fewest disguising records purchased, achieving the highest utility. In general, for both attacks, as $\alpha$ gradually increases from 0 to 1, indicating a higher allocation of weight to the total number of records in the published intent rather than to the gain in satisfying the privacy constraint during expansion, the number of disguising records decreases. This corresponds to a gradual increase in utility.

\section{Conclusions}\label{sec:con}

In this article, we tackle an interesting and important problem in data markets--preserving data buyers' privacy. We formulate the problem and presents a disciplined approach exploring the trade-off between privacy and data purchase cost. The thorough experimental results clearly show the effectiveness and efficiency of our approach.

Our approach can be extended and generalized to tackle more sophisticated cases, such as complicated data purchase intent beyond a conjunctive normal form and incorporating cost factors into the utility computation. As future work, it is also interesting to explore individual buyers' privacy cost and the distinct characteristics of data buyers versus general product buyers.

\section*{Acknowledgement}

This work was supported in part by a Beyond the Horizon grant by Duke University.  All opinions, findings, conclusions and recommendations in this paper are those of the authors and do not necessarily reflect the views of the funding agencies.

\section{}



\ifCLASSOPTIONcaptionsoff
  \newpage
\fi



%



\bibliographystyle{IEEEtran}
\bibliography{references}

%








\end{document}